\documentstyle[11pt,epsf,fleqn]{article}
\newcommand{\Del}{$\Delta$}

\newcommand{\thalf}{$\frac{3}{2}$ }

\def\beq{\begin{equation}}
\def\eeq{\end{equation}}
\def\bea{\begin{eqnarray}}
\def\eea{\end{eqnarray}}
\def\eqref#1{Eq.~(\ref{eq:#1})}
\def\eqlab#1{\label{eq:#1}}

\def\tabref#1{Table \ref{tab:#1}}

\def\NP#1#2#3{Nucl. Phys. {\bf #1} (#2) #3}
\def\PRL#1#2#3{Phys.~Lett. {\bf #1} (#2) #3}
\def\PR#1#2#3{Phys.~Rev.~{\bf #1} (#2) #3}
\def\PRL#1#2#3{Phys.~Rev.~Lett. {\bf #1} (#2) #3}
\def\AP#1#2#3{Ann. of Phys. {\bf #1} (#2) #3}
\def\VYP#1#2#3{{\bf #1} (#2) #3}  

\def\half{\mbox{\small{$\frac{1}{2}$}}}

\begin{document}
\voffset-1.0cm
\textheight 22cm
\textwidth 15cm
\thispagestyle{empty}
\hfill k-matr.tex
\vspace{20pt}

\begin{center}
{\LARGE
Pion and photon induced reactions on the nucleon in a unitary model.}\\
\vspace{20pt}

{\large O. Scholten$^{a)}$, A. Yu. Korchin$^{a,b)}$, V. Pascalutsa$^{a,c)}$
and  D. Van Neck$^{a,d)}$}\\
\vspace{7pt}
{\small\it $^{a)}$Kernfysisch Versneller Instituut, 9747 AA Groningen, The
Netherlands. \\[0.1cm]
 $^{b)}$National Scientific Center Kharkov Institute of Physics and
Technology, 310108 Kharkov,
     Ukraine$^1$. \\[0.1cm]
 $^{c)}$Institute for Theoretical Physics, Utrecht University, 3508 TA Utrecht,
 The Netherlands$^2$.  \\[0.1cm]
 $^{d)}$University of Gent, Proeftuinstraat 86, B-9000 Gent, Belgium$^2$.
} \\[0.2cm]
\end{center}
{\small   $^1$Permanent address.  \\
 $^2$Present address.  }
\vspace{2cm}

\centerline{\bf Abstract}
\begin{quote}

     We present a relativistic calculation of pion scattering, pion
photoproduction  and Compton scattering on the nucleon in the energy region
of the
\Del-resonance (upto 450 MeV photon lab energy), in a unified framework which
obeys the unitarity constraint.   It is found that the recent data on the cross
section for nucleon Compton scattering determine accurately the parameters of
the electromagnetic nucleon--\Del\ coupling. The calculated
pion-photoproduction partial-wave amplitudes  agree well with the recent Arndt
analysis.
\end{quote}
\bigskip

\vfill
\hfill  \today


\section{Introduction}

    It is well known that in Compton scattering from the nucleon, by using
arguments based on unitarity and causality \cite{Gel54}, strong constraints can
be put on the cross section. These are usually formulated in terms of
the fixed-t dispersion relations \cite{disp}  expressing the real part of
the six Hearn-Leader amplitudes \cite{Hea62} through the imaginary part. The
latter is directly related by the optical theorem to  the pion-photoproduction
cross section. In such an approach the relatively small Born terms are added to
account for the low-energy behaviour of the full amplitude. The slow
convergence of the dispersion integrals requires several subtraction functions
which are related to the t-channel singularities, especially to the $\pi\pi$
exchange \cite{Pfe74}. Since the latter is poorly determined one has to use
some models for the multiple meson exchange, thus making the approach model
dependent.  A lower unitary bound on the Compton cross section can be obtained
by setting the real parts of the amplitude equal to zero \cite{disp} or to
the Born
contribution \cite{Cap82}. This of course allows to avoid the uncertainties of
the dispersion approach, however the results obtained seem to agree with
the data in
the \Del-resonance region thus leaving no room for the proper real parts
calculated from the dispersion integral.  On the other hand in this approach
the link with the decay properties of the nucleon resonances is getting
obscured. In particular it is difficult to extract the contribution of the
\Del-resonance to Compton scattering and parameters of the decay $\Delta \to
N+\gamma$.

    In order to express more clearly the contribution of the \Del-resonance, a
relativistic tree-level calculation was performed\cite{Pas95} for nucleon
Compton scattering. In this calculation the \Del-resonance was included via the
Rarita-Schwinger propagator with a complex self energy to account for its pion
decay width and thus, implicitely, for the pion channels. The calculation
showed that even at energies near the pion threshold, $E_{\gamma}\approx 150$
MeV, the contribution of the \Del-resonance is crucial. Even though a good
agreement with the data was obtained, one aspect missing in the calculation is
that unitarity is obeyed only approximately.

    From calculations on pion photoproduction (see, e.g.,\cite{Dav91,VdH95}) it
is known that the unitarity constraint, which can for example be imposed via
Watson's theorem connecting $(\pi ,\pi)$ and $(\gamma ,\pi)$ amplitudes, is
crucial to obtain the correct interference between the resonant and the
background contributions. In this work we have improved on the calculation of
ref.\cite{Pas95} in this respect.

\section{Outline of the model}

 A simple approach that is particularly suited for imposing the unitarity
constraint while keeping at the
same time a direct link to the basic Feynman diagrams, is the $K$-matrix
approach. In this approach the T-matrix, $S=1+2iT$, is represented as $T=
K/(1-iK)$, from which it is evident that the scattering matrix $S$ is unitary
when the $K$-matrix is hermitian.

 We have employed the $K$-matrix approach in the space $\pi N \oplus
\gamma N$. In this way Compton scattering is investigated together with the
pion-nucleon scattering and pion photoproduction. This has the
additional advantage that all three processes are calculated {\it consistently}
 which puts stronger constraints on the model parameters.

    Due to time-reversal invariance the partial wave $K$-matrix will be a real
and symmetric $4\times 4$ matrix in a basis spanned by two pion-nucleon
channels, corresponding to different values for the total $\pi N$ isospin (1/2
and 3/2) and two photon-nucleon channels, corresponding to different helicities
(or, equivalently electric and magnetic radiation).
For the partial wave decomposition  we use the Jacob--Wick formalism as given
in appendices of refs. \cite{Pfe74,Ber67}.

  Within the model space unitarity
is satisfied exactly, however the two pion channel, which is known to become
important at energies in excess of 400 MeV, is not included.
We find that though the phases in the partial wave amplitudes
for Compton scattering off the proton have changed considerably as compared to
the non-unitarized calculation\cite{Pas95}, the cross sections have changed
little in the
energy region upto the \Del -resonance. At the resonance and beyond there are
substantial differences.

   The $K$-matrix is approximated by the sum of tree-level diagrams including
direct ($s$-type) and crossed ($u$-type) 'driving' terms with intermediate
nucleon, $N^{\star}$ (Roper)- and \Del -resonances with  {\it real} self
energies equal to the mass of the resonances.
In the $t$-channel for pion scattering, $\sigma$- and $\rho$-meson exchanges
are taken into account. In pion photoproduction the $\pi$-, $\rho$- and
$\omega$-meson are included in the t-channel, where the latter two have only a
marginal effect on the calculated quantities and the first is necessary to
ensure current conservation. In Compton scattering only $\pi^0$-meson exchange
is included.

   The $K$-matrix formalism results from the Bethe-Salpeter
equation in the approximation that the principal value of the loop integrals
is neglected
and only the contribution from the discontinuity is kept. Stated differently,
the particles forming loops are taken to be on the mass shell.
The width of the resonances is generated
dynamically in the calculation of the $T$-matrix as a result of iteration of
the direct diagrams. The pole contributions from the loops involving $u$- and
$t$-type  diagrams in the $K$-matrix give rise to the pion-loop vertex
corrections to the $\pi NN$ and $\gamma NN$ vertex functions.  Both the decay
width and vertex corrections are thus generated dynamically in the $K$-matrix
approach. Because of this internal consistency the unitarity constraint is
exactly satisfied. Two-pion channel is not included in the model space. Since
the phase-space for this channel opens only gradually it can safely be ignored
up to photon energies of about 400 MeV.

   As mentioned only the pole contributions from the pion (and photon) loops
are included and, for reasons of simplicity, the principle-value part of
the full 4-dimensional integral
is neglected. The effect of the latter has been studied in the framework of a
relativistic integral equation in which the pion is restricted to its mass
shell in Ref. \cite{Gro93} for the $\pi N \to \pi N$ and in Ref. \cite{Sur95}
for $\gamma N \to \pi N$ reactions. It can however be argued \cite{Gou94} that
the effect of the real part of the loop integrals will be mainly a
renormalization of the coupling constants and the masses of the particles
involved. In our calculations these are taken to be the physical values where
known and are otherwise treated as free parameters; we believe that such
renormalizations are probably of small importance.

    The \Del-state is included via the spin-\thalf\ Rarita-Schwinger
propagator,  which off-shell contains also a spin-\half\ background. In each of
the vertices involving the \Del\ therefore also an off-shell coupling parameter
enters, which determines the coupling to the spin-\half\ sector of the
Rarita-Schwinger propagator. These off-shell couplings  appear to be of crucial
importance to reproduce the data. In the $\pi N N$ coupling vertex we have
allowed for a mixture of pseudo-scalar and pseudo-vector coupling specified by
parameter $\chi$ ($\chi=0$ corresponds to pure pseudo-vector coupling). For
non-zero values of $\chi$ the Kroll-Rudermann term is included in pion
photoproduction to restore gauge invariance.

\section{Results and Discussion.}

   As emphasized, in the calculations the unitarity constraint is satisfied,
even to higher orders in $\alpha$, the fine structure constant. For $\pi$N
scattering the higher order corrections in $\alpha$ are negligible and we will
therefore discuss this case first to fix the pion coupling parameters.

   Our model for $\pi$N scattering is very similar to that of
Goudsmit et al.\cite{Gou94}.  The main difference is that, since we are
interested in somewhat higher energies, we have also included the
Roper-resonance in the calculations.
This improves the fit in the $P_{11}$ channel as
to be expected, but hardly influences any of the other partial-wave amplitudes.
To investigate the effect of the strong couplings we have
used  the results of three different fits to the $\pi$N phase shifts. Two of
these, parameter sets \# 1 and \# 2 in \tabref{t1}, have been taken from
ref.\cite{Gou94}. These two parameter sets allow for the investigation of the
effect of pseudo-scalar  v.s.\ pseudo-vector $\pi$NN-coupling. To study the
effect
of the off-shell coupling in the $\pi$N\Del vertex (characterized by
the parameter $z_{\pi}$) we have analysed
also the third set given in \tabref{t1}.

\begin{table}[htb]
\begin{center}
\begin{tabular}{c|ccccccc}
\hline \hline
set \# & $g_{\pi N N}$ & $\chi$ & $g_{\pi N \Delta}$ & $z_{\pi}$ &
       $G_{\sigma}$ & $g_{\rho N N}$ &   $\kappa_{\rho}$ \\
\hline
  1 & 12.95  & 0.0 & 2.19 & -.34 &  23 & 5.87   &   2.1 \\
  2 & 12.95  & 0.2 & 2.19 & -.34 &  43 & 2.90   &   2.1 \\
  3 & 12.95  & 0.0 & 2.19 & -.16 &  28 & 5.40   &   2.1 \\
 \hline \hline
\end{tabular}
\caption[T1]{  \label{tab:t1}
{\small Different sets of parameters used in the calculation of the
pion-nucleon scattering. In the definition of the interaction Lagrangian ref.\
\cite{Gou94} is followed, only the PV/PS mixing parameter $x$ is renamed to
$\chi$. Parameter sets \# 1 and \# 2 correspond to two fits to the $\pi N$
scattering
data as presented in ref.\ \cite{Gou94}, at the extremes of their parameter
spectrum. All parameters not explicitely mentioned are taken from this work
with $g_{\pi \pi \rho} = 6.065$. The Roper resonance is included following
ref.\ \cite{Gar93} with $H=0.145$ and vanishing width. Only its one-pion
partial decay width is generated dynamically. } }
\end{center}
\end{table}

    All three parameter sets given in \tabref{t1} give a comparable overall fit
to the Arndt partial-wave data\cite{Arn95}. Parameter set \# 3 gives somewhat
better results at higher energies in the $S_{11}$-channel but slightly worse
results in the $S_{31}$-channel. Since for the first two parameter sets our
results
are very similar to those of ref.\cite{Gou94} we will not discuss these any
further in this short letter.

\begin{table}[htb]
\begin{center}
\begin{tabular}{c|ccccc}
\hline \hline
set \# &  $G_1$   &   $z_1$   &   $G_2$   &   $z_2$  & $R(E2/M1)$ \\
\hline
  1   &   4.3   &   0.15  &   2.0   &   4.0  & -4.47\%\\
  2   &   4.3   &   0.05  &   4.0   &   2.5  & -2.56\%\\
  3   &   4.3   &   0.0   &   6.0   &   1.8  & -0.57\%\\
 \hline \hline
\end{tabular}
\caption[T2]{\label{tab:t2}{\small
Different sets of parameters for the $\gamma N \Delta$ vertex that give a
comparable fit to the cross--section data. The parameters for the $\omega$
meson couplings are taken from ref.\ \cite{Gar93}. For the electro-magnetic
decay of the Roper resonance we used\cite{Gar93} $G_p=-0.544$ and
$G_n=+0.552$. The $E2/M1$ decay ratio is defined according to \eqref{Rem}.}}
\end{center}
\end{table}

   The results of our calculations for the pion photoproduction partial-wave
amplitudes and the Compton-scattering cross section are compared with the data
in Figs.\  1 and 2. In these calculations only the four parameters of the
$\gamma$N\Del vertex have been optimized.  As shown in \tabref{t2}, there is a
range of values for $G_2$, compensated by an appropriate change in $z_1$ and
$z_2$, for which a comparably good fit can be obtained. A best fit to the data
for Compton scattering is obtained with parameter set \# 1 from \tabref{t2}.

A parameter often quoted for the N$\Delta\gamma$-vertex is the E2/M1 ratio
for the electromagnetic decay of the \Del-resonance. This ratio
is however not directly related to physical observables due to
background contributions. It is defined as the ratio of the electric and
magnetic decay rate of an 'on-shell' \Del-resonance. As such it depends only
on the parameters $G_1$ and $G_2$ and {\em not} on the off-shell coupling
parameters
$z_1$ and $z_2$\cite{Dav91},
\beq
R({E2\over M1})=-{2\,G_1-G_2 {M_{\Delta} \over M} \over
          2\,G_1{3M_{\Delta}+M \over M_{\Delta}-M}-G_2 {M_{\Delta} \over M} }
\eqlab{Rem}
\eeq
where $M$\ ($M_{\Delta}$) is the nucleon (\Del-resonance) mass.
The predictions for R, as
given in \tabref{t2}, thus vary strongly with $G_2$ while keeping agreement
with the data where the variation of $G_2$ is compensated with a variation
of the off-shell parameters.

In pion photoproduction the partial wave amplitudes are largely insensitive to
the off-shell parameters.
From Figs.\  1 and 2 one can see that the different choices for $G_{2}$ affect
only the $E_{1+}^{3/2}$ multipole amplitude
\footnote{The notation ${\cal M}_{L\pm}^{I}$ is used in pion
photoproduction where ${\cal M}$ stands for the electric (${\cal M}=E$) or
magnetic (${\cal M}=M$) type of the photon with the final $\pi N$ state
characterized by the orbital angular momentum $L$, parity $p=(-1)^{L+1}$,
total angular momentum $J=L
\pm 1/2$ and total isospin $I=1/2,\, 3/2$. In pion scattering $L_{2I\,2J}$
is used. In Compton scattering the notation
is $f_{\cal M\,M'}^{L\pm}$ where the total angular momentum is $J=L
\pm 1/2$ and  parity is $p=(-1)^{L}$ for ${\cal M}=E$ or $p=(-1)^{L+1}$ for
${\cal M}=M$.}
in pion photoproduction at energies exceeding 300 MeV.
 None of the other partial wave amplitudes are affected.

 In Compton scattering a variation of $G_{2}$ mainly affects the $f_{EE}^{2-}$
and the $f_{ME}^{1+}$ amplitudes which are related to the $E_{1+}^{3/2}$
amplitude in the pion photoproduction channel.
The $\gamma N\Delta$ vertex enters
the amplitude quadratically and the strong vertices do no enter explicitely.
Therefore Compton scattering is strongly dependent  on the off-shell
coupling to the background spin-\half\ fields. The corresponding
off-shell coupling parameters are thus determined rather accurately. This
observation goes in line with the finding \cite{Pea91,Gou94} for the
pion-nucleon
scattering where a strong dependence on the off-shell parameter $z_\pi$
in the $\pi N N$ vertex has been obtained.

   None of the different choices for the parameters for $\pi$N scattering
given in \tabref{t1} do noticeably affect the results for Compton scattering.
Including a pseudo-scalar coupling improves the agreement with the
$\pi$-N-scattering data at higher energies but has hardly any influence on the
pion photoproduction channel. There is a considerable  sensitivity to the
off-shell coupling parameter in the $\pi$N\Del vertex, $z_{\pi}$. As mentioned,
the fit for the $S_{31}$ $\pi$N phase shift is somewhat worse for parameter set
\# 3 in \tabref{t1}, but the agreement for pion photoproduction is considerably
improved, especially for the $M_{1+}^{1/2}$ channel. Again, the Compton
scattering calculations are not affected.

     For energies well below the \Del-resonance the unitarity constraint is not
very important when calculating the Compton cross section. This is shown in
Fig.\ 3 where the results of the present calculations are compared with the
non-unitarized tree approximation where a finite width for the \Del-resonance
is included\cite{Pas95}. Only at energies where the \Del-resonance is
dominating the spectrum the differences are appreciable.

   In this figure also the result of a calculation is shown where the
photon-decay of the \Del\ is switched off. This shows clearly the importance of
the \Del\ even at energies near the pion threshold. Only due to a strong
destructive interference at small angles and a strong constructive interference
at large angles of the nucleon- and the \Del-amplitudes the characteristic
vanishing of the cross section at 0 degrees can be accounted for.

\bigskip

   In conclusion,
Compton scattering is demonstrated to be most suitable for
determining the parameters of the $\gamma N \Delta$ vertex since the Compton
amplitude is largely {\it insensitive} to the strong channels, while it is very
{\it sensitive} to the photon-\Del\ coupling. Quite the opposite is found for
 pion photoproduction.
We have shown that in a rigorously unitary calculation of
Compton scattering from the proton  the interference of the nucleon and
\Del-isobar amplitudes is crucial to understand the observed structure in the
cross section as was also found in a tree level calculation\cite{Pas95}.
The unitary constraint turns out to be important in the calculation of
the cross section at photon energies exceeding 250 MeV.

\bigskip

One of us (A.Yu.\ K.) thanks the staff of the KVI for the kind hospitality
extended to him at the KVI Groningen. He would also like to thank the
Netherlands Organization for International Cooperation in Higher Education
(Nuffic) and the
Nederlandse organisatie voor Wetenschappelijk Onderzoek (NWO) for financial
support during his stay in the Netherlands.

     This work was financially supported by de Stichting voor Fundamenteel
Onderzoek der Materie (FOM).


\begin{itemize}

\item[fig. 1] {The pion-photoproduction partial-wave amplitudes are compared
with the analysis of ref.\cite{Arn90}. In particular it is shown that the
effect of changing $G_2$ (\tabref{t1}, set \# 1 and \tabref{t2}, set \# 3) is
small as compared to the norm calculation (\tabref{t1}, set \# 1 and
\tabref{t2}, set \# 1). The off-shell parameter in the $N\Delta\gamma$-vertex,
$z_{\pi}$, is much larger as shown in the third calculation using (\tabref{t1},
set \# 3 and \tabref{t2}, set \# 1).}

\item[fig. 2] {A comparison of different calculations for Compton scattering
the parameters used are the same as for figure 1. The data for the
Compton-scattering cross section are from ref.\ \cite{Hal93,Bla96}}

\item[fig. 3] {The results of the full ('Norm' calculation of figures 1 and 2)
calculation for the Compton cross section is compared with a calculation in
which the coupling of the photon to the \Del-resonance is set to zero ('No
Delta') and with a calculation along the lines of ref.\cite{Pas95} using the
same values for the parameters in the photon coupling vertices ('No
Unitarity').}
\end{itemize}

\end{document}